\documentclass[twocolumn]{article}

\usepackage{amssymb,amsmath,graphicx}

\title{Stability of Information in the Heat Flow Clustering}
\author{Brian Weber}
\date{December 18, 2023}

\begin{document}
	\maketitle
	
	\begin{abstract}
		Clustering methods must be tailored to the dataset it operates on, as there is no objective or universal definition of ``cluster,'' but nevertheless arbitrariness in the clustering method must be minimized.
		This paper develops a quantitative ``stability'' method of determining clusters, where stable or persistent clustering signals are used to indicate real structures have been identified in the underlying dataset.
		This method is based on modulating clustering methods by controlling a parameter---through a thermodynamic analogy, the modulation parameter is considered ``time'' and the evolving clustering methodologies can be considered a ``heat flow.''
		When the information entropy of the heat flow is stable over a wide range of times---either globally or in the local sense which we define---we interpret this stability as an indication that essential features of the data have been found, and create clusters on this basis.
	\end{abstract}	
	
	\section{Introduction}
	Clustering is essential in many computing an data analysis applications, but it is long understood that some apriori characterization of what should count as a cluster \cite{Castro02} must be made before deciding on an appropriate method.
	In many cases, particularly when noise is serious enough, different clustering algorithms can give widely different results.
	Some algorithms require initial selections, for example an initial choice of centers, and even small differences in such choices can lead to serious divergences in outcomes.
	Appropriate choice of methods or initial parameters often require foreknowledge of what kind of structures to expect within the dataset, and practitioners sometimes supply boutique or ad hoc rules based on such foreknowledge, or just use the researcher's intuition.
	In this paper we present a method that substantially reduces the arbitrariness involved in cluster searches: \emph{information stability} in a time-dependent clustering method that we call \emph{heat-flow clustering}.
	
	We take a dataset to be a collection of $N$ many points $X=\{{\bf{x}}_l\}_{l=1}^N\subset\mathbb{R}^n$, and a \emph{clustering} of $X$ to be a partition $\{X_i\}_{i=1}^M$ of $X$ into $M$ many subsets, meaning a collection of subsets $X_i=\{{\bf{x}}_{i,l}\}_{l=1}^{N_i}$ for which $X_i\cap{}X_j=\varnothing$ when $i\ne{}j$, and $\bigcup_i{}X_i=X$.
	We define the extropy of the clustering $\{X_j\}$ to be the partition's Shannon measure of information (SMI) \cite{Shannon48}, which is
	\begin{equation}
		S(\{X_i\})\;=\;-\sum_{i=1}^M\frac{N_i}{N}\,\log\,\frac{N_i}{N} \label{EqnSMIDef}
	\end{equation}
	where $N_i$ is the cardinality of $X_i$ and $N=\sum_iN_i$ is the cardinality of the dataset $X$.
	
	We use potential-theoretic methods to exemplify the heat-flow technique.
	Potential methods have the notable advantage of not needing to know the number of clusters ahead of time, but a serious disadvantage in the question of how to choose a kernel.
	The potential or data-field clustering that appeared in \cite{FWJ10} begins with choosing a kernel $k:\mathbb{R}^n\rightarrow\mathbb{R}^{+}$ with $\int{}k({\bf{x}})dVol_n=1$, and convolving with the datapoints of $X$ to obtain a potential or ``data-field''
	\begin{equation}
		P({\bf{x}})\;=\;\frac{1}{N}\sum_{i=1}^Nk\left({\bf{x}}-{\bf{x}}_i\right).
	\end{equation}
	The local maxima of $P$ are taken to be cluster centers.
	Clustering can then be performed using methods native to potential-theoretic methods, such as the level-set clustering of \cite{WGLL11} or differencing-potential (DP) method of \cite{WW18}, or else using a secondary clustering method such as K-means.
	In Sections \ref{SecOneD} and \ref{SecHigherD} we create our own easy-to-use potential-based clustering method.
	We emphasize that the clustering technique itself is actually not so important; what is needed is a way of modulating the clustering method and measuring stability.
	Potential-based methods make this easy because a kernel, a Gaussian kernel in particular, is easily adjusted in a one-dimensional fashion by altering its variance.
	
	Indeed a serious problem in potential methods is choosing the variance of the kernel: too small and every datapoint becomes a cluster, too large and essential features of the data are blended together, even to the point of reducing everything to just a single cluster.
	Arranging these kernels in order of increasing variance is (up to possible reparameterization) the same thing as performing a heat flow.
	See Fig.~\ref{EqnHeatFlowFigure} for a depiction.
	To be more specific we parabolically rescale the kernel by setting
	\begin{equation}
		K({\bf{x}},t)
		\;=\;\frac{1}{t^n}\,k\left({\bf{x}}/t\right)  \label{EqnTimeVaryingKernel}
	\end{equation}
	so we retain $\int{}K({\bf{x}},t)\,dVol_n\equiv1$ but encounter the spreading of the kernel's standard deviation.
	When $k({\bf{x}})=e^{-\frac14\|{\bf{x}}\|^2}$ is the Gaussian then $K=K({\bf{x}},t)$ solves the parabolic equation $2\frac{1}{t}K_t=\triangle{}K$.
	After the reparemterization $t=\sqrt{2s}$, $K$ solves the standard heat equation $K_s=\triangle{}K$.
	The reason for choosing the parameterization of (\ref{EqnTimeVaryingKernel}) rather than the standard heat-flow parameterization, is to obtain a spreading of the kernel's standard deviation linearly with time, rather than by the square root of time, which would distort our stability analysis.
	
	Creating an ordered selection $t_0<t_1<\dots<t_T$ of ``time'' values, for each $t_k$ we obtain a potential
	\begin{equation}
		P({\bf{x}},\,t_k)
		\;=\;
		\frac{1}{N}\sum_{l=1}^NK({\bf{x}}-{\bf{x}}_l,\,t_k) \label{EqnTimeKernel}
	\end{equation}
	and from this potential a clustering $\{X_{k,i}\}_{i=1}^{M_k}$ of $X$ at time $t_k$.
	At each time $t_k$ we record two pieces of information: the number of clusters $M_k$, and the SMI
	\begin{equation}
		S_k=S(\{X_{k,i}\}_{i=1}^{M_k})
		=-\sum_i\frac{N_{k,i}}{N}\log\frac{N_{k,i}}{N} \label{EqnTimeEntropy}
	\end{equation}
	where $N_{k,i}$ is the cardinality $N_{k,i}=\#X_{k,i}$.
	Below, we create a way of measuring the \emph{stability over time} of these measures, and create a clustering that optimizes this stability.
	
	Intuitively, we interpret stability as an indication that real, underlying features of the dataset have been detected.
	Incidental features and noise, by contrast, would be expected to produce ephemeral effects that might affect the clustering for certain values of $t$, but disappear for most others.
	
	Referring to the modulation parameter as ``time'' is largely for convenience, but there is an analogy with physical processes.
	In aggregate processes such as heat dissipation, the fact that the system's entropy increases with time coincides with a loss of information as the aggregate system moves from more ``surprising'' or improbable states to more ``expected'' or probable states.
	Whether it models a physical process or not, the entropy of the potential function $P({\bf{x}},t)$ increases with time, even if non-Gaussian kernels are used or the time steps are disuniform; see for example \cite{Day93}, \cite{Evans04}, or \cite{Ni04} for interpretations of entropy for continuous flows.
	The point is that even though $P$ might or might not not solve an actual heat equation, its behavior through time is qualitatively similar to a heat flow.
	
	To stretch the thermodynamic analogy a bit further, we are interested in a certain \emph{state function}: the cluster partition, which through (\ref{EqnSMIDef}) carries its own entropy.
	The increasing of the entropy of the heat potential $P$ coincides with the decreasing of the entropy in the clustering.
	If the potential and the clustering are taken together as a whole system, as time progresses the internal information in the potential comes to dominate, washing out the information carried by the cluster partition.
	This observation, on the time-dependence of the information entropy, is the basic rationale behind calling this the heat-flow clustering.
	
	
	%

	\section{The chronodendrogram and information stability} \label{SecChronoStab}
	
	The parameter $t$, responsible for modulating the clustering method, allows creation of a hierarchical clustering with layers that depend on time.
	After creating linkages between clusters at adjacent time slices, the result is a graph we call the \emph{chronodendrogram}, or time-dependent tree, associated to the dataset.
	
	Let $X=\{{\bf{x}}_i\}_{i=1}^N\subset\mathbb{R}^n$ be a dataset, choose time values $t_0<t_1<\dots<t_T$, and let $\{X_{k,i}\}_{i=1}^{N_k}$ be the clustering at time $t_k$.
	To each cluster $X_{k,i}$ we associate a node, and to create linkages between nodes we join the cluster $X_{k,i}$ at time $t_k$ to the cluster $X_{k-1,i'}$ at time $t_{k-1}$ with weight $w_{k,i,i'}=\#(X_{k,i}\cap{}X_{k-1,i'})$.
	This is the \textit{inheritance} linkage: the weight on the edge connecting two clusters is the number of points inherited from the earlier cluster to the later cluster.
	For examples see Figs.~\ref{EqnOneDChronoDendro}, \ref{FigOneDComplicatedDendrogram}, and \ref{Fig2DChronoDendro}.
	
	To measure stability, at each time $t_k$ we assign two global variables: the number of clusters $M_k$, and the entropy $S_k$.
	Each $M_k$ is a positive integer, and to each $n\in\mathbb{N}$ we define the \textit{stability score}
	\begin{equation}
		B(n)\;=\;\frac{\#\{k\;|\;M_k=n\}}{T+1}.
	\end{equation}
	$B(n)$ is simply the proportion of times $t_k\in\{t_0,\dots,t_T\}$ for which the number of clusters is $n$.
	
	The score $B(n)$ alone is insufficient to identify stable clusters, because although the number of clusters might be stable over time, the clusters might be exchanging datapoints.
	This phenomenon is captured by the entropy (\ref{EqnTimeEntropy}).
	If the number of clusters is constant, the entropy is constant if and only if the number of datapoints in each cluster is also constant.
	(This still leaves the possibility of clusters exchanging points while keeping the totals constant, but in practice this is uncommon.)
	Given a number of clusters $n$ and an entropy range from $s_1$ to $s_2$, we define the associated \emph{entropy stability score} to be
	\begin{equation}
		B_{s_1}^{s_2}(n)\;=\;\frac{\#\{k\;|\;M_k=n\;\text{and}\;s_1<S_k\le{}s_2\}}{T+1}.
	\end{equation}
	This measures the time-range for which the dataset is divided stably into $n$ many clusters within a specified entropy range.
	However, a dataset will often have very stable individual clusters even if the dataset as a whole is less stable.
	To capture this phenomenon we create the notion of \emph{local stability}.
	Let $X'\subseteq{}X$ be any subset; normally this will be a cluster at some intermediate time $t_k$, although this is not strictly necessarily.
	We define the \emph{backtrack} of $X'$ from some time $t_k$ to be the following clustering at all times earlier that $t_k$:
	\begin{equation}
		X'_{k',i}\;=\;X'\cap{}X_{k',i}.
	\end{equation}
	Simply put, the backtrack $\{X_{k',i}\}_{i=1}^{M'_{k'}}$ is a clustering of $X'$ created by intersecting $X'$ with the already-existing clustering of the larger set $X$.
	Similar to the global numbers $M_k$ and $S_k$ associated to $X$, we have local numbers $M'_{k'}$ and $S'_{k'}$ where $M'_{k'}$ is the number of non-empty sets $X'_{k'}$ in the partition of $X'$ at time $t_{k'}$ and the entropy at time $t_{k'}$, $S'_{k'}=S(\{X'_{k',i}\}_{i=1}^{M_k'})$.
	We define the \emph{local entropy stability score} of $X'$ at time $t_k$ to be
	\begin{equation}
		\begin{aligned}
		&B_{s_1}^{s_2}(X',t_k)\\
		&=\frac{\#\{k'\;\big|\;\text{$k'\le{}k$, $s_1<S'_{k'}\le{}s_2$}\}}{T+1}.
		\end{aligned}
	\end{equation}
	In Sections \ref{SecOneD} and \ref{SecHigherD} we work through examples that use this local score to identify highly stable clusters within datasets that display lower stability overall.
	
	A useful, but certainly not the only, algorithm for identifying stable clusters is the following.
	Find the diameter of the dataset; the time at which the kernel $k$ has standard deviation about half this diameter is when the potential $P$ should no longer be able to identify any internal clusters, but is expected to find only a single cluster in the dataset.
	Let this be the upper bound time $t_T$ (larger standard deviations will only give the same result, that everything is in one cluster).
	Choose a lower bound $t_0$ be the time value that gives the standard deviation of $k$ twice the pixel dimension (smaller standard deviations will only create one-pixel clusters, which is useless).
	Depending on the datset, this consolidation into one cluster may occur sooner and it is useful to let $t_T$ be the earliest time this occurs; we call this the \emph{consolidation time}.
	After creating a uniform partition of the interval from $t_0$ to $t_T$, $t_0<t_1<\dots<t_{T-1}<t_T$, measure $B(n)$ for all $n$ between $1$ and $N$ and find the value of $n=n'$ that maximizes $B(n)$.
	Choose $T'$ to be the largest integer so the number of clusters $M_{T'}$ at time $t_{T'}$ is $B(n')$.
	Then at $t_{T'}$ create a local analysis by letting $X'$ be any of the clusters observed at time $t_{T'}$, and determining the stability of each.
	Clusters that meet an appropriate threshold can be removed from the dataset, and the process repeated for the remaining clusters, until all points are clustered.

	\section{One dimensional clustering} \label{SecOneD}
\begin{figure}[h]
		\includegraphics[scale=0.2]{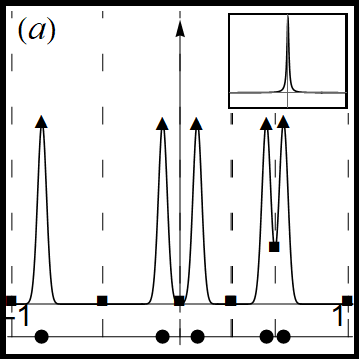}
		\includegraphics[scale=0.2]{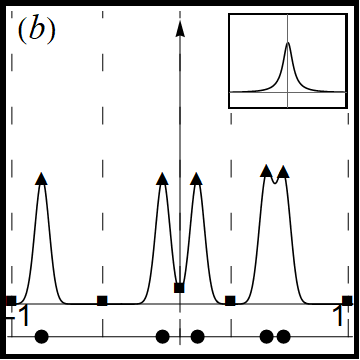}
		\includegraphics[scale=0.2]{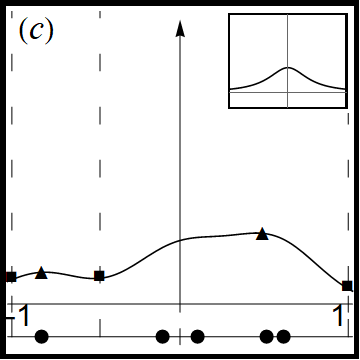}
		\caption{
			Three time slices for the heat flow clustering of the dataset (\ref{EqnToyModelData}).
			Datapoints (circles along the lower axis) are depicted with maxima (triangles) and minima (squares) of the potentials.
			Cluster selections are indicated by dashed lines.
			Inset is the kernel choice.
			In the first subfigure we find five clusters, in the second three, and in the third two.
		} \label{EqnHeatFlowFigure}
	\end{figure}

\begin{figure}[h!]
		\includegraphics[scale=0.23]{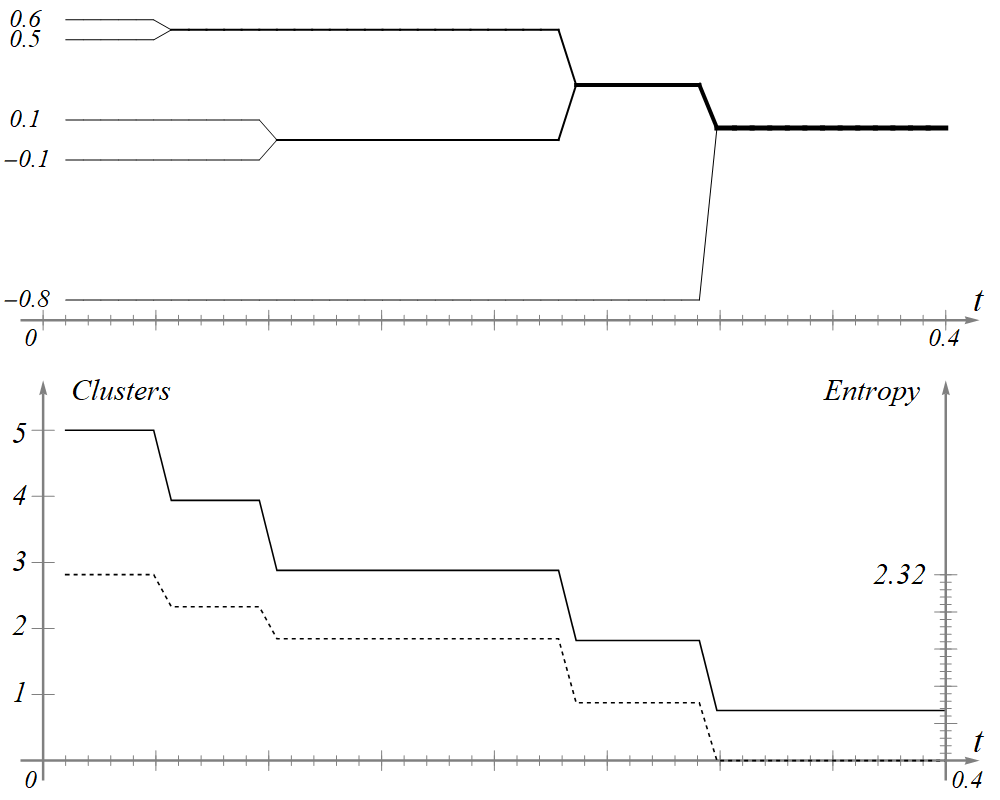}
		\caption{
			Above: The chronodendrogram for the dataset with five elements; linkage weights are indicated by thickness.
			Below: The two informational measures, $M_k$ and $S_k$, as a function of time.
		} \label{EqnOneDChronoDendro}
	\end{figure}

	We give two examples of our heat-flow stability clustering on one-dimensional datasets.
	The first uses a simple dataset of just the five elements
	\begin{equation}
		X=\{-0.8,\,0.0,\,0.2,\,0.5,\,0.6\}. \label{EqnToyModelData}
	\end{equation}
	See Figs.~\ref{EqnHeatFlowFigure} and \ref{EqnOneDChronoDendro}.
	We start with such a simple example because it is easily tractable but still shows most of the essential features of the method.
	We choose 50 evenly spaced time values between $t_0=0.01$ and $t_{20}=0.4$.
	For each time value we create the potential $P$, and perform the \emph{minimum-partition clustering}, as follows.
	Let $\mu_0,\dots,\mu_K$ be the list of local minima of $P$ at time $t_k$.
	Then $\mathbb{R}^1$ partitions into segments $L_i=[\mu_{i-1},\mu_i)$, $1\le{}i\le{}K$ and $L_0=(-\infty,\mu_0)$ and $L_{K+1}=[\mu_K,\infty)$.
	For each $t_k$ we create the partition $\{L_{k,i}\}_{i=1}^{K_k}$ using the minima of $P(x,t_k)$, and then partition $X$ by setting
	\begin{equation}
		X_{k,i}\;=\;X\cap{}L_{k,i}.
	\end{equation}
	With this clustering method we obtain the chronodendrogram for the dataset (\ref{EqnToyModelData}) depicted in Fig.~\ref{EqnOneDChronoDendro}.
	
	In the selected time range we have the following persistence scores: $B(5)=B(4)\approx0.117$, $B(3)\approx0.333$, $B(2)\approx0.157$, and $B(1)\approx0.27$.
	The most stable case is therefore the case of three clusters, which are $X_1=\{0.6,0.5\}$, $X_2=\{0.1,-0.1\}$, and $X_3=\{-0.8\}$.
	The clusters consolidate into one at time $t=0.3$.
	If we take the consolidation time $t=0.3$ to be the terminal time $t_T$, as indicated in the algorithm of Section \ref{SecChronoStab}, we find an even clearer signal:
	\begin{equation}
		\begin{aligned}
		&B(5)=B(4)\approx0.161, \\
		&B(3)\approx0.460,\;B(2)\approx0.217.
		\end{aligned}
	\end{equation}
	
	\begin{figure}[h]
		\hspace{0.05in}
		\includegraphics[scale=0.25]{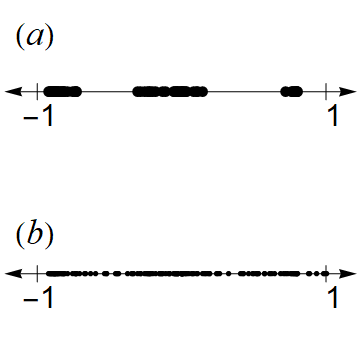} \hspace{0.05in}
		\includegraphics[scale=0.25]{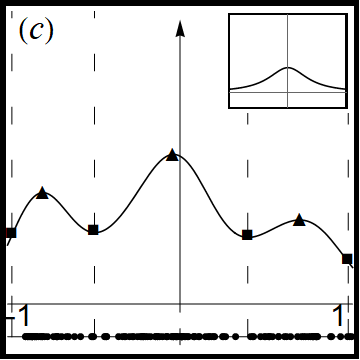}
		\caption{
			(a) displays the three clusters with varying densities and numbers of points, (b) displays the full dataset with random noise added, and (c)  shows the potential at $t=0.158$, the time indicated in Fig.~\ref{FigOneDComplicatedDendrogram}.
		} \label{FigOneDComplicated}
	\end{figure}
\begin{figure}[h]
		\includegraphics[scale=0.23]{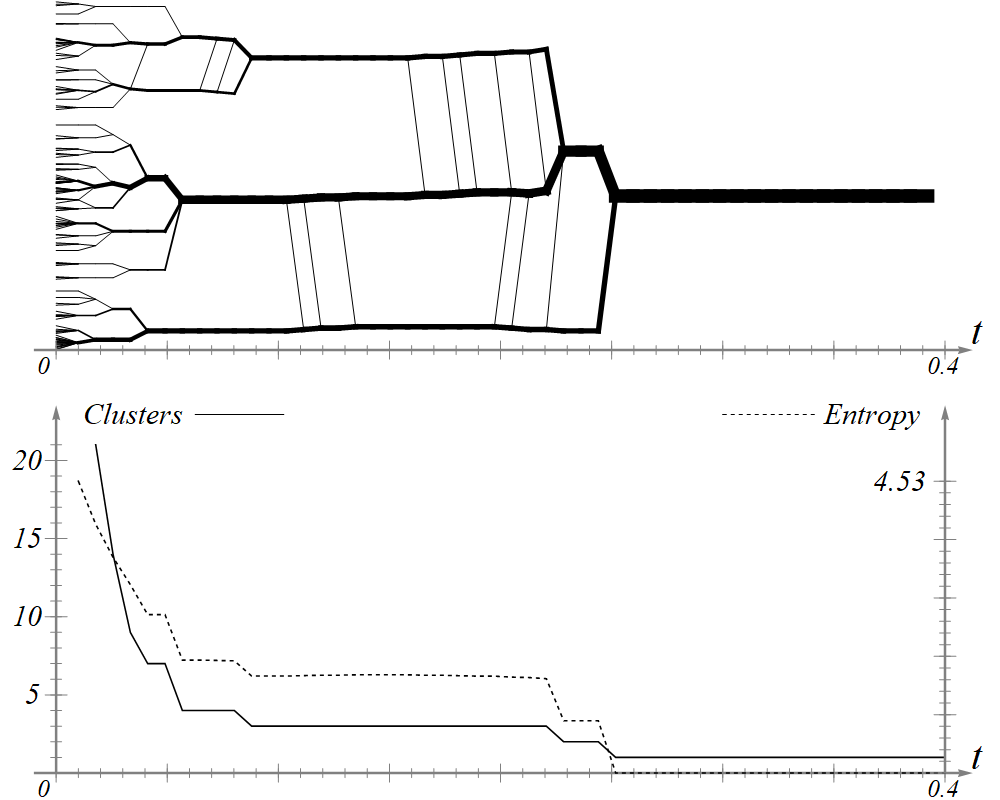}
		\includegraphics[scale=0.268]{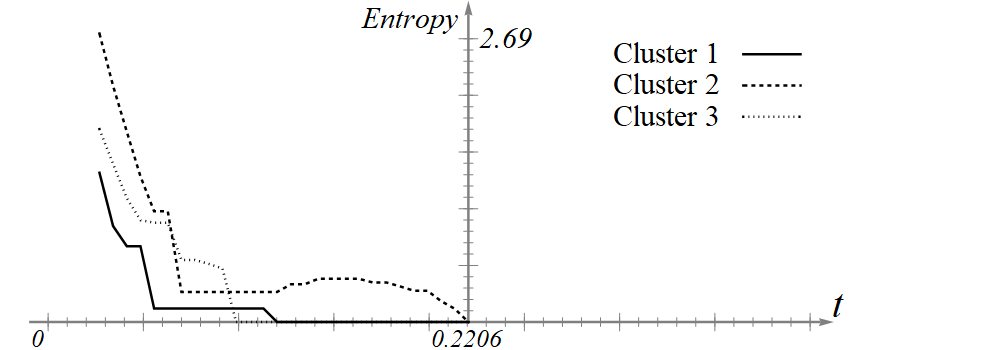}
		\caption{
			Top: Chronodendrogram for the noisy dataset.
			Middle: The entropy and cluster number, showing stability at 3 clusters.
			Bottom: The local entropy scores for each of the three clusters observed at time $t=0.2206$.
		} \label{FigOneDComplicatedDendrogram}
	\end{figure}

	For our second example we create a larger dataset with three clusters of different diameters and densities, and a large amount of noise.
	To create the clusters we picked 30 random points between $-0.9$ and $-0.7$, 30 random points between $-0.3$ and $0.2$, and 10 random points between $0.7$ and $0.8$.
	In addition to these 70 intentionally clustered points, we choose 70 random points between $-1$ and $1$ to simulate noise.
	See Figs. \ref{FigOneDComplicated} and \ref{FigOneDComplicatedDendrogram}.
	The global stability numbers are
	\begin{equation}
		\begin{aligned}
		&B(4)\approx0.078,\;
		B(3)\approx0.353, \\
		&B(2)\approx0.059,\;
		B(1)\approx0.392,
		\end{aligned}
	\end{equation}
	and $B(n)$ for $n\ge5$ is uniformly less than $0.04$.
	This gives a very strong stability measure of $35\%$ for $3$ clusters.
	The consolidation time is somewhat earlier than in the previous example, at about $t=0.26$.
	Using $t_T=0.26$ rather than $t_T=0.4$,
	\begin{equation}
		\begin{aligned}
			&B(4)\approx0.108,
			B(3)\approx0.487,
			B(2)\approx0.0813.
		\end{aligned}
	\end{equation}
	Despite the very strong signal of $0.487$ for three clusters, the global analysis hides an even stronger signal within the dataset.
	Time $t_k=0.2206$ at $k=28$ is the largest time-value for which we have three clusters, $X_{k,1}$, $X_{k,2}$, $X_{k,3}$.
	Localizing to the first cluster $X'=X_{k,1}$ and using the consolidation time $t_T=0.26$ as the terminal time,
	\begin{equation}
		B_{0.0}^{0.1}(X',0.2206)\;=\;0.86
	\end{equation}
	indicating the entropy of this cluster measured at $t=0.2206$ remains extremely small, between $0.0$ and $0.10$, for $86\%$ of the entire time interval from $t=0$ to $t=0.26$.
	See Fig.~\ref{FigOneDComplicatedDendrogram}.
	Therefore cluster 3, as measured at time $t=0.2206$, shows greater stability than any of the other clusters measured in the dataset.
	As shown in Fig.\ref{FigOneDComplicatedDendrogram}, clusters 1 and 2 at time $t=0.2206$ are also reasonably stable, but over shorter time intervals and with larger entropy variations.

	\section{Higher dimensions} \label{SecHigherD}
	Higher dimensional data can displays wider patterns of behavior.
	First we create a means of clustering high dimensional data.
	
	Let $X=\{{\bf{x}}_i\}_{i=1}^N\subset\mathbb{R}^n$ be a dataset.
	With $K({\bf{x}},t)=\frac{1}{\pi{}t^2}e^{-|{\bf{x}}|/t^2}$, create the potential $P({\bf{x}},t)=\frac{1}{N}\sum_{i=1}^NK({\bf{x}}-{\bf{x}}_i,t)$.
	Choosing times $\{t_0,\dots,t_T\}$ we find a list of local maxima $\{{\bf{m}}_{k,j}\}_j$, which we consider to be the cluster centers at time $t_k$.
	The assignment of points to cluster centers must proceed differently than in the one-dimensional case, because in higher dimensions the minima of a function do not partition the configuration space.
	Given any datapoint ${\bf{x}}\in{}X$ and any local maximum ${\bf{m}}={\bf{m}}_{k,j}$, we define a cost function $Cost({\bf{x}},{\bf{m}})$ in the following way.
	Let $\gamma:[0,1]\rightarrow\mathbb{R}^2$ be the segment
	\begin{equation}
		\gamma(s)=(1-s){\bf{x}}\,+\,s{\bf{m}}
	\end{equation}
	from the datapoint ${\bf{x}}$ to the maximum ${\bf{m}}$.
	Letting $P:\mathbb{R}^n\rightarrow\mathbb{R}$ be the potential function (suppressing its time-dependency for the moment), we define the cost to be the normalized total variation of the potential along this path:
	\begin{equation}
		\begin{aligned}
		&Cost({\bf{x}},{\bf{m}}) \\
		&\quad=\frac{1}{|P({\bf{x}})-P({\bf{m}})|}
		\int_0^1\left|\frac{d}{ds}P\circ\gamma\right|\,ds.
		\end{aligned} \label{EqnCost}
	\end{equation}
	This cost is bounded from below by $1$, and reaches this theoretical minimum if and only if $P$ is monotonic along $\gamma$ (and therefore monotonically increasing; it cannot be monotonically decreasing because the endpoint ${\bf{m}}$ is a local maximum).
	It is possible the cost reaches this minimum for several choices of ${\bf{m}}$; in that case we select among these the nearest to ${\bf{x}}$.
	(In the unlikely case that both cost and distance are equal among several choices of local maxima, we assign ${\bf{x}}$ to one of these maxima at random.)
	\begin{figure}[h]
		\hspace{0.2in}\includegraphics[scale=0.2]{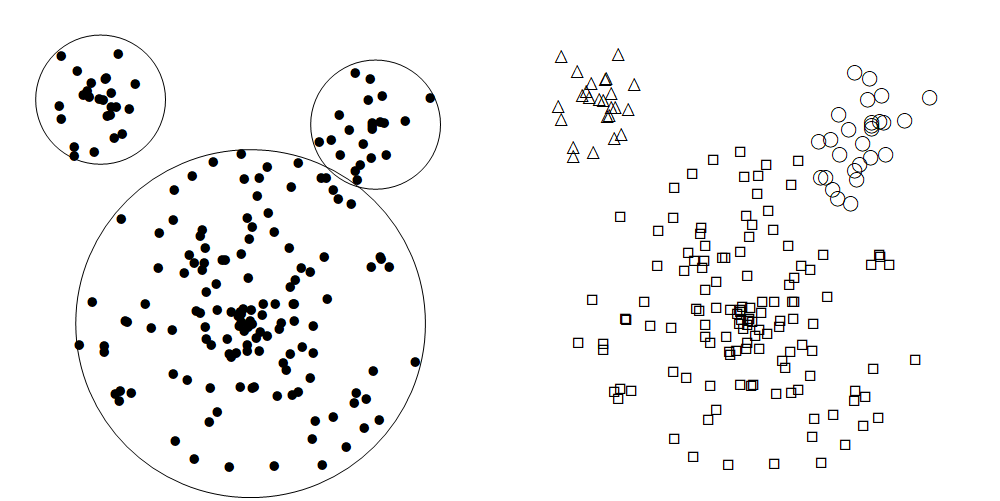}
		\caption{
			Left: The data points, distributed randomly within three circles.
			Right: Clustering actually obtained at time $t_k=0.1833$.
		} \label{Fig2DPts}
	\end{figure}
	\begin{figure}[h]
		\includegraphics[scale=0.225]{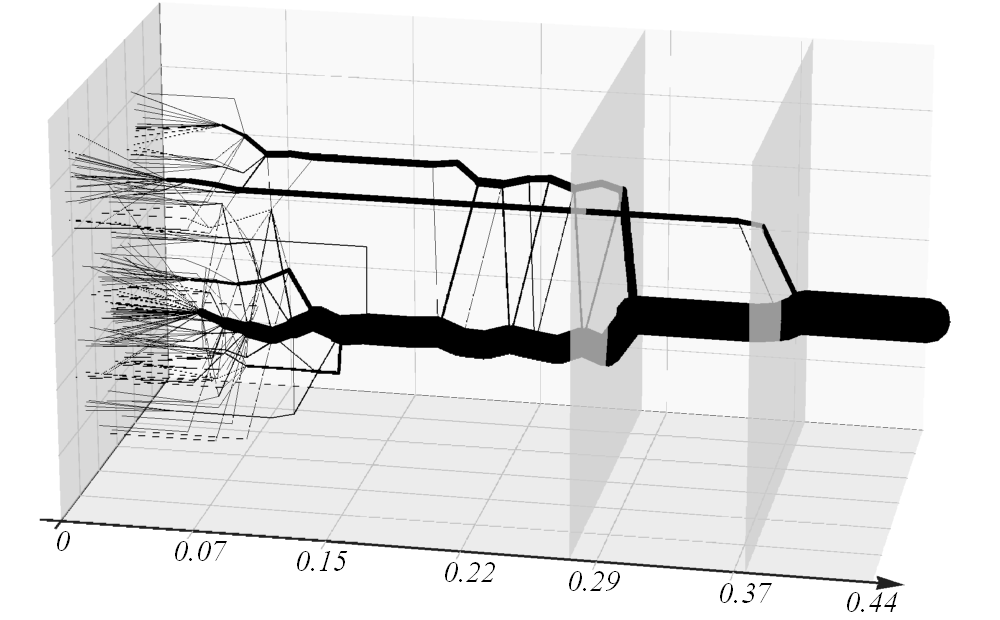}
		\caption{
			The chronodendrogram for the dataset of Fig.~\ref{Fig2DPts}.
			Slices at $t_k=0.277$ and $t_k=0.377$ indicated.
		} \label{Fig2DChronoDendro}
	\end{figure}
	\begin{figure}[h!]
		\includegraphics[scale=0.225]{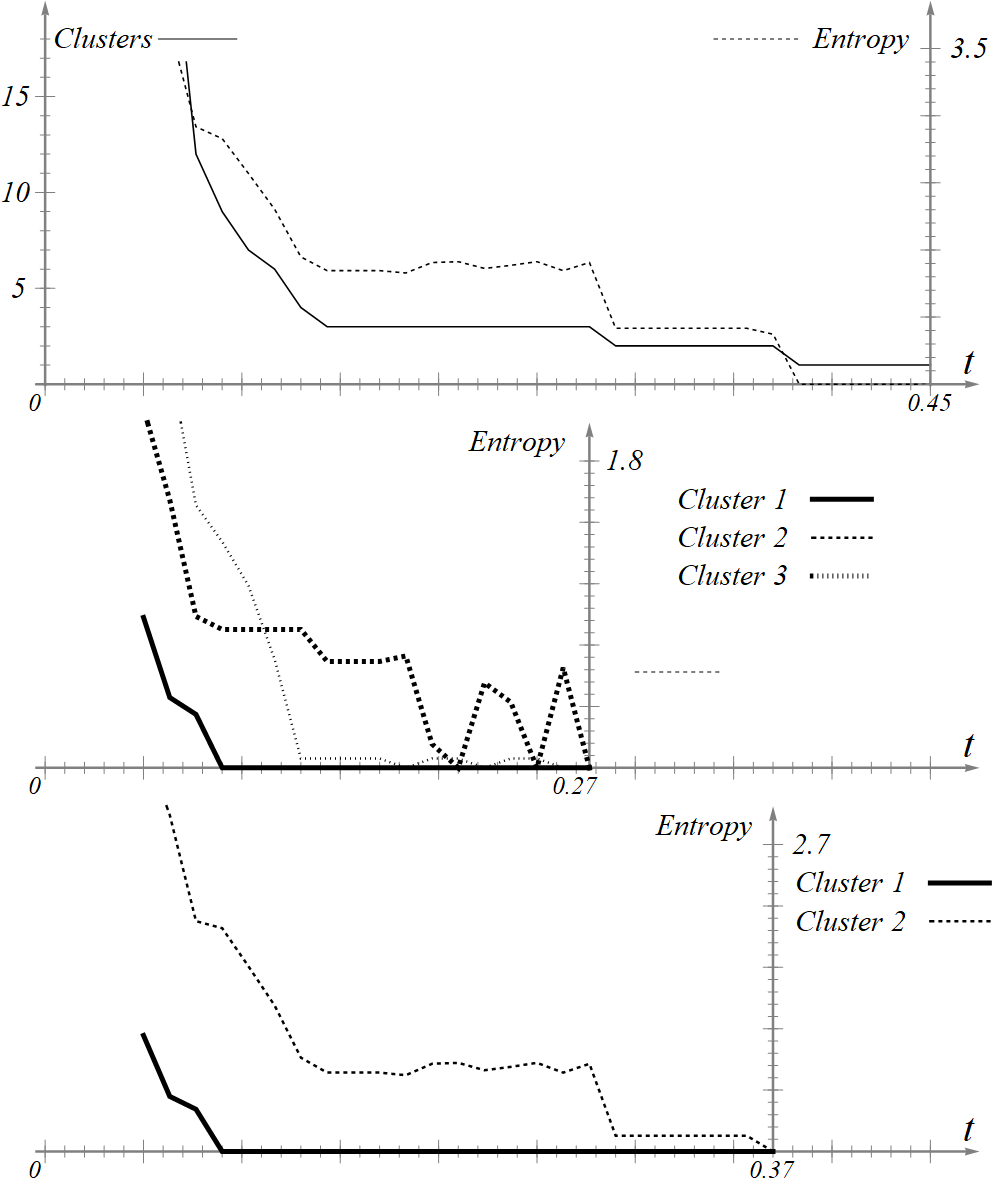}
		\caption{
			Top: Number of clusters and entropy.
			Middle: Local entropy analysis for the three clusters observed at $t_{18}\approx0.28$, showing notable stability in Clusters 1 and 2, but substantial instability in Cluster 3.
			Bottom: Local entropy scores for the two clusters observed at $t_{25}\approx0.37$ showing a very long period of zero entropy for Cluster 1.
		} \label{Fig2DEntropy}
	\end{figure}
	
	As in the introduction, we remark that many other forms of clustering are available, for example a K-means with the local maxima ${\bf{m}}_{k,j}$ as initial cluster centers, or the DP method of \cite{WW18}.
	
	We consider the heat-flow clustering analysis on the dataset in Fig.~\ref{Fig2DPts}.
	This dataset $X\subset\mathbb{R}^2$ is created by selecting random points within three different circles.
	In each of the two smaller circles are 25 randomly placed points, and in the larger circle is 150 randomly placed points.
	For the heat flow we use a Gaussian kernel with 31 evenly spaced time slices $t_k$ between $t_0=0.05$ and $t_{30}=0.45$, and perform the clustering, with the method above based on the cost function (\ref{EqnCost}).
	The resulting chronodendrogram is shown in Fig.~\ref{Fig2DChronoDendro} and the global entropy measure, along with the cluster number and entropy measures, is shown in Fig.~\ref{Fig2DEntropy}.
	We find
	\begin{equation}
		B(3)=0.355,\;
		B(2)=0.236,\;
		B(1)=0.194,
	\end{equation}
	and $B(n)$ is uniformly smaller than $0.04$ for $n>3$.
	At $B(3)\approx0.36$ the case of $3$ clusters is most probable; however global entropy measure fluctuates within $1.214\pm0.059$, a range of about $10.0\%$ around the median, indicating non-trivial instabilities.
	This appears in Fig.~\ref{Fig2DChronoDendro} as substantial interchange between two of the branches.
	
	We take two local measures of stability, one at time $t_k=0.277$ and one at time $t_k=0.37$, as indicated by the slices in Fig.~\ref{Fig2DChronoDendro}.
	The entropy measures for $t_k=0.18$ is indicated in Fig.\ref{Fig2DEntropy}.
	At $t_k=0.277$, which is $k=18$, we observe three clusters $X_{18,1}$, $X_{18,2}$, $X_{18,3}$.
	The first two clusters are unstable, but letting $X'_1=X_{18,1}$, $X'_2=X_{18,2}$, $X'_3=X_{18,3}$ we find we the local entropy
	\begin{equation}
		\begin{aligned}
		B_{0.0}^{0.0}(X'_3,0.277)\;=\;0.415. \label{Eqn2DLocalPeristence}
		\end{aligned}
	\end{equation}
	Cluster 3 then has high persistence with zero entropy.
	The persistence measure (\ref{Eqn2DLocalPeristence}) is based on the maximum time value $t_T=0.45$, but selecting instead $t_T=0.383$, the point at which all clusters accumulate into one, we find the stronger signal $B_{0.0}^{0.0}(X'_3,0.277)=0.487$.
	
	In fact this cluster is even more stable than the local analysis at $t_{18}\approx0.277$ indicates.
	Doing local analysis at the larger value $t_k=0.377$, the second slice in Fig.~\ref{Fig2DChronoDendro}, we find two clusters.
	The local entropy at $0.377$ is depicted in Fig.~\ref{Fig2DEntropy}.
	The smaller of them has zero entropy over a very long time; we have
	\begin{equation}
		B_{0.0}^{0.0}(X',0.377)\;=\;0.687
	\end{equation}
	indicating this cluster persists, completely unchanged, for $67.7\%$ of the time interval from $t_0=0.05$ to $t_T=0.45$, lending a high degree of confidence to this cluster.
	Using the consolidation time $t_T=0.383$ as the largest time value, we find the even larger stability score of $B_{0.0}^{0.0}(X',0.377)=0.795$ for this cluster.

	\section{Discussion}
	
	The heat-flow clustering method described here was developed to solve a specific problem in the labeling of raw datasets coming out of laboratory scans of electron configurations in semiconductor-based quantum computing devices \cite{WZ24}.
	There, it was found that the usual clustering methods such as K-means and nearest-neighbor were unstable enough that end results needed to be checked by human experts, negating the purpose of automating the process in the first place.
	Investigating sources of this instability, it was found in apriori choices were at the root and making ``good'' apriori choices depended on information that varied widely among datasets such as how many clusters exist, how diffuse or concentrated individual clusters might be, and how near or far cluster centers might be from one another.
	All of this was complicated by high levels of obscuring noise which caused too much blending of clusters in many methods.
	
	Potential-based methods, for the right choice of kernel, created a low-pass filtering that eliminated most of the noise, but if chosen badly could also blend together clusters.
	While it was difficult to make specific rules for an apriori choice of kernel, stability was observed over wide ranges of choices, albeit over different ranges in different datasets.
	Choosing the clusters on the basis of observed stability lead to a fully automatic clustering method that succeeded on every dataset it was tested on.
	
	Before closing we mention that the work \cite{WGLL11} proposed a potential-based clustering method with an automated way of choosing a specific kernel.
	That proposal was to choose a kernel whose variance minimizes the entropy measure
	\begin{equation}
		S_{WGLL}\;=\;-\sum_i\frac{P({\bf{x}}_i)}{Z}\log\frac{P({\bf{x}}_i)}{Z} \label{EqnWGLLEntropy}
	\end{equation}
	where $Z=\sum_iP({\bf{x}}_i)$ (see Eq. (8) of \cite{WGLL11}, or Chapter 3.2 of \cite{Fritz97} for a similar notion).
	Lacking another name, we call this the Wang-Gan-Li-Li entropy.
	However in practice this minimum is often unsuitable.
	For example in many reasonable datsets there are multiple local minima of $S_{WGLL}$, in others the minimum is unusably close to $t=0$.
	This minimum is unstable with respect to even modest variations in the underlying dataset; it is most stable when clusters are about the same size and similar distances apart, but is incapable of detecting clusters that occur at different scales from one another, something our local analysis is well suited for.

\end{document}